\documentclass[11pt]{article}
\usepackage{amsmath,amssymb}

\def\ZZ{\mathbb Z} 
\def\CC{\mathbb C}
\def\NN{\mathbb N} 
\def\RR{\mathbb R}
\def\SS{\mathbb{S}}
\def\HH{\mathcal H} 
\def\AA{\mathcal A}
\def\FF{\mathcal F}
\def\BB{\mathcal B}

\def\OO{\mathrm{O}}
\def\SO{\mathrm{SO}}
\def\UU{\mathrm{U}}
\def\SU{\mathrm{SU}}
\def\Mp{\mathrm{Mp}}
\def\Sp{\mathrm{Sp}}

\def\so{\mathfrak{so}}
\def\uu{\mathfrak{u}}
\def\su{\mathfrak{su}}

\def\sp{\mathfrak{sp}}
\def\gg{\mathfrak{g}}
\def\kk{\mathfrak{k}}

\def\ul{\underline}
\def\ket#1{\vert\,\ul #1\,\rangle}
\def\bra#1{\langle\,\ul #1\,\vert}
\def\braket#1#2{\langle\,\ul #1\,\vert\,\ul #2\,\rangle}
\def\vac{\vert0\rangle} \def\cav{\langle0\vert}
\def\eps{\varepsilon}

\def \M {\overline{M}}

\def \z {\mathrm{z}}
\def \w {\mathrm{w}}
\def \u {\mathrm{u}}
\def \v {\mathrm{v}}
\def \har {\mathfrak{h}}
\def \Res {\mathrm{Res}}

\def\bea{\begin{eqnarray}} \def\eea{\end{eqnarray}}

\begin{document}
%%%%%%%%%%%%%%%%%%%%%%%%%%%%%%%%%%%%%%%%%%%%%%%%%%%%%%%%%%%%%%%%%%%%%%%%%%%%
\renewcommand{\today}{}
\renewcommand{\theequation}{\thesection.\arabic{equation}}
\title{\bf Unitary Positive-Energy Representations \\ 
of Scalar Bilocal Quantum Fields}
\author{Bojko Bakalov$^{1^*}$, Nikolay M.\ Nikolov$^{2,3^*}$, \\[2mm]
  Karl-Henning Rehren$^{3^*}$, Ivan Todorov$^{2,3^*}$}
\date{
\bigskip
\begin{itemize}
\item[$^{1}$] {\small Department of Mathematics, North Carolina State
    University,} \\ {\small Box 8205, Raleigh, NC 27695, USA}
\item[$^{2}$] {\small Institute for Nuclear Research and Nuclear Energy,} \\
{\small Tsarigradsko Chaussee 72, BG-1784 Sofia, Bulgaria}
\item[$^{3}$] {\small Institut f\"ur Theoretische Physik,
    Universit\"at G\"ottingen,} \\ {\small Friedrich-Hund-Platz 1,
    D-37077 G\"ottingen, Germany}
\end{itemize}
}

\maketitle

\begin{abstract}
The superselection sectors of two classes of scalar bilocal quantum
fields in $D\geq 4$ dimensions are explicitly determined by working out the
constraints imposed by unitarity. The resulting classification in terms
of the dual of the respective gauge groups $\UU(N)$ and $\OO(N)$ confirms
the expectations based on general results obtained in the framework of
local nets in algebraic quantum field theory, but the approach using
standard Lie algebra methods rather than abstract duality theory is
complementary. The result indicates that one does not lose interesting
models if one postulates the absence of scalar fields of dimension
$D-2$ in models with global conformal invariance. Another remarkable
outcome is the observation that, with an appropriate choice of the
Hamiltonian, a Lie algebra embedded into the associative algebra of
observables completely fixes the representation theory.

\end{abstract}

\vskip5mm

PACS 2003: 11.10.--z, 03.70.+k

MSC 2000: 17B65, 22E65, 81T10, 81R10

\vskip6mm

\begin{itemize}
\item[$^{*}$] {\small e-mail addresses: 
{\tt \par bojko{\char95}bakalov@ncsu.edu \\
mitov@inrne.bas.bg, nikolov@theorie.physik.uni-goe.de \\
rehren@theorie.physik.uni-goe.de  \\
todorov@inrne.bas.bg, itodorov@theorie.physik.uni-goe.de}}
\end{itemize}
\newpage
%\tableofcontents

\section{Introduction}
\label{sec1}\setcounter{equation}{0}

An important tool in the study of {\em globally conformal invariant}
(GCI) quantum field theory models in even space-time dimensions $D\geq 4$
\cite{NT01,NST02,NST03,NRT05} are bilocal fields, which arise in 
operator product expansions (OPE) as follows. 

Let $\phi(x)$ be a local scalar field of dimension $d\geq 2d_0 := D-2$.
We denote the contribution of twist $D-2$ in the OPE of 
$((x_1-x_2)^2)^{d-d_0}\cdot \phi^*(x_1)\phi(x_2)$ by
\bea\label{W}
W(x_1,x_2)=W(x_2,x_1)^*
\eea
if $\phi(x)$ is {\em complex}, and by 
\bea\label{V}
V(x_1,x_2)=V(x_2,x_1) = V(x_1,x_2)^*
\eea
if $\phi(x)=\phi^*(x)$ is
{\em real}. This means that the expansion of $W(x_1,x_2)$ into local
fields yields an infinite series of conserved symmetric traceless
tensor currents, starting with the scalar field $W(x,x)$ of
dimension $2d_0$ (which may be zero), and including the 
stress-energy tensor. The similar expansion of $V(x_1,x_2)$ contains
only tensors of even rank. It is a nontrivial consequence of GCI
that the twist $D-2$ fields $W$ and $V$ are {\em bilocal} in the sense
of (strong = Huygens) local commutativity with respect to both
arguments \cite{NRT06}.  

Depending on the scaling dimension $d$ of the field $\phi(x)$
whose OPE produces the bilocal field, the latter may exhibit
different singularities in its correlation functions. In the case 
$d=D-2$ when the singularities have the lowest possible degree, one can
derive the commutation relations of the bilocal field, and they
involve another bilocal field with the same properties. 
If we assume uniqueness of the bilocal field, then it can be
normalized in such a way that the commutation relations take the form
\bea \label{ccomm}
[W(x_1,x_2),W(x_3,x_4)] = \Delta_{2,3} \; W(x_1,x_4) +
\Delta_{1,4} \;W(x_3,x_2) + N\cdot \Delta_{12,34}
\eea
in the complex case, and
\bea \label{rcomm}
[V(x_1,x_2),V(x_3,x_4)] = \Delta_{2,3} \; V(x_1,x_4) +  \Delta_{2,4}
\; V(x_1,x_3) + \qquad\qquad\nonumber \\ + \Delta_{1,4} \; V(x_2,x_3) +
\Delta_{1,3} \;V(x_2,x_4) + N\cdot (\Delta_{12,34} +
\Delta_{12,43})
\eea
in the real case, where $\Delta^+_{i,j}=\Delta_+(x_i-x_j)$ and
$\Delta_{i,j}=\Delta^+_{i,j}-\Delta^+_{j,i}$ are the two-point and
commutator functions of a massless free scalar field (which has
dimension $d_0$), and $\Delta_{12,34} = \Delta^+_{1,4}\Delta^+_{2,3} - 
\Delta^+_{4,1}\Delta^+_{3,2}$. The coefficients $N$ are the normalizations
of the four-point functions 
\bea \label{4point}
&& \cav W(x_1,x_2)W(x_3,x_4)\vac = N \cdot
\Delta^+_{1,4}\Delta^+_{2,3} , \nonumber \\
&& \cav V(x_1,x_2)V(x_3,x_4)\vac
= N \cdot (\Delta^+_{1,4}\Delta^+_{2,3} + \Delta^+_{1,3}\Delta^+_{2,4}).
\eea
The above relations are such that if $W(x_1,x_2)$ satisfies (\ref{ccomm}), 
then $V(x_1,x_2)=W(x_1,x_2)+W(x_2,x_1)$ 
satisfies (\ref{rcomm}) with $2N$ instead of $N$.

Our goal in this paper is to explore and rule out the possibility of
realizations of the above commutator relations in GCI models, other
than the free field realizations
\bea \label{Wfree}
W(x_1,x_2) = {:}\vec\varphi^{\, *}(x_1)\cdot\vec\varphi(x_2){:} =
\sum_{p=1}^N{:}\varphi^{p*}(x_1)\varphi^p(x_2){:}
\qquad\hbox{(complex)}
\eea
where $\varphi^p(x)$ are $N$ mutually commuting complex massless free
fields of dimensiond $d_0$, and
\bea \label{Vfree}
V(x_1,x_2) = {:}\vec\varphi(x_1)\cdot\vec\varphi(x_2){:} =
\sum_{p=1}^N{:}\varphi^p(x_1)\varphi^p(x_2){:} \qquad\hbox{(real)}
\eea
where $\varphi^p(x)$ are $N$ mutually commuting real massless free
fields. (Free-field constructions of bilocal fields involving spinor
or vector fields can also be given; their correlations
exhibit higher singularities.)

Clearly, the free field realizations exist only when $N$ is a positive
integer. Indeed, it was shown in \cite{NST02} for the real
case that due to Hilbert space positivity in the vacuum sector the
coefficient $N$ in (\ref{rcomm}) must be a positive integer 
and the bilocal field $V(x_1,x_2)$ is of the form (\ref{Vfree})
(or $N=0$ corresponding to the trivial case $V=0$). 
As a byproduct of our considerations, we establish
this result of \cite{NST02} both in the complex and the real case.

In particular we deduce that, if $\phi(x)$ is a real scalar
field  of dimension $D-2$, which is assumed to be the unique such
field in the model, and $V(x_1,x_2)$ arises in the OPE of $\phi(x)$ with
itself, then $V(x,x)$ is a multiple of $\phi(x)$ and thus $\phi(x)$
is a sum of Wick squares. Dropping the uniqueness assumption, a
similar statement is still true, but then a linear combination of
bilocal fields of the form (\ref{Vfree}) with different positive
coefficients may appear \cite{NRT06}. 

It is important to observe that expressions (\ref{Wfree}) and (\ref{Vfree}) are
defined only on the Fock space of the free fields or on subspaces
thereof. When discussing theories containing a scalar field
of dimension $D-2$ one has to envisage the possibility of different
representations of the associated bilocal field occurring in the full
Hilbert space $\HH$.

A pertinent general structure theorem in the framework of algebraic QFT
\cite{DR90} states that all superselection sectors of a local QFT
$\AA$ are contained in the vacuum representation of a canonically
associated (graded local) field extension $\FF$, and they are in a
one-to-one correspondence with the irreducible unitary representations
of a compact gauge group $G$ (of the first kind) of internal
symmetries of $\FF$, so that $\AA\subset \FF$ consists of the fixed
points under $G$. The gauge group is unitarily implemented in
the vacuum Hilbert space $\HH$ of $\FF$, and the central projections
in $U(G)''$ decompose this Hilbert space into inequivalent
representations of $\AA\subset U(G)'$. \footnote{$U(G)=\{U(g):g\in G\}$ 
are the unitary representers of the gauge
  group, $U(G)'$ is the commutant of $U(G)$, i.e., the algebra of
  bounded operators on $\HH$ which commute with every element of
  $U(G)$, and $U(G)''$ is the commutant of $U(G)'$, which coincides
  with the weak closure of the linear span of $U(G)$.} 

In the above free field representation, the bilocal fields are
the fixed points under the gauge group $\UU(N)$ or $\OO(N)$ in the
theory of $N$ free complex or real massless scalar fields,
respectively. Since the latter is known to have no nontrivial
superselection sectors (\cite[Sect.\ 3.4.5--6]{R90} and \cite[App.\
A]{BDLR92}), one can conclude that it coincides with the above
canonical field net $\FF$, 
so that the sectors of the bilocal field are in correspondence with
the representations of $G=\UU(N)$ or $\OO(N)$. However, the general
theorem of \cite{DR90} evokes a great amount of abstract group
duality, and its application requires the passage from fields to nets
of local algebras and back. Although this passage is well understood
in the case of free fields, it would be desirable to see the
comparatively simple assertion emerge by more elementary methods.    

We are interested in the \emph{unitary positive-energy} representations 
of the infinite dimensional Lie algebras of local commutators (\ref{ccomm})
and (\ref{rcomm}). We shall pursue two alternative formulations of the
problem. The first consists in defining energy positivity with respect
to the conformal Hamiltonian canonically expressed in terms of the
fields (Eq.~(\ref{Hcan}) below). In this case, we do not assume in advance
the free field realizations (\ref{Wfree}) and (\ref{Vfree}) in the vacuum
sector, and not even that $N\in\NN$ in the commutation relations;
these properties will be deduced instead from unitarity (cf.\ \cite{NST02}).

In the second formulation, in order to deal with a more general class
of ``additively renormalized'' Hamiltonians (see the next section for
details), given the free field realizations in the vacuum sector we
assume that the representations are generated from the vacuum by
(relatively) local fields. This allows us to use the Reeh--Schlieder
theorem \cite{RS61}, according to which every local relation among
Wightman fields which holds on the vacuum vector must hold in the full
Hilbert space $\HH$ generated by other (relatively) local
fields. Apart from the commutation relations, there are polynomial
relations of the form 
\bea \label{pol}
\det \Big(W(x_i,x_j)\Big)_{i,j=1}^{N+1} - \hbox{contraction terms} =0,
\eea
which arise from (\ref{Wfree}) and (\ref{Vfree}) by expanding the
left-hand-side of the identity 
\bea \label{detwick}
{:}\left[\det 
\Big(\vec\varphi^{\, *}(x_i)\cdot\vec\varphi(x_j)\Big)_{i,j=1}^{N+1}
\right]{:} = 0
\eea
as a polynomial in the normal products 
${:}\vec\varphi^{\,*}(x_i)\cdot\vec\varphi(x_j){:}$ 
in accord with Wick's theorem (and similarly in the real case).
The determinant relations (\ref{pol}) are valid in the Fock space
representation, and hence, by the Reeh--Schlieder theorem, also in
other superselection sectors. Therefore, the algebra of bilocal fields
may be regarded as the quotient of the associative enveloping algebra
of the Lie algebra (\ref{ccomm}) or (\ref{rcomm}) by the determinant
relations (\ref{pol}) and possibly some additional
relations\footnote{Our result shows that 
  if additional relations exist, then they are satisfied automatically
  in every unitary positive-energy representation satisfying the
  determinant relations.}.    

We shall demonstrate that in both formulations of the problem, one
arrives at the same classification of unitary positive-energy
representations (superselection sectors) of the Lie algebras
(\ref{ccomm}), (\ref{rcomm}) of bilocal fields. We find precisely
those representations that occur in the Fock space of $N$ free scalar
fields, and they are in a one-to-one correspondence with the
irreducible unitary representations of $\UU(N)$ in the complex case or
$\OO(N)$ in the real case. 

A possible application to the program of classification of GCI 
models is the following. If bilocal fields satisfying (\ref{ccomm}) or
(\ref{rcomm}) appear in the OPE of some local fields of the model,
then all other (relatively) local fields, which possibly intertwine
different superselection sectors, may be regarded as Wick products of
a free field multiplet transforming in some representation of $\UU(N)$
or $\OO(N)$, possibly tensored with some other GCI fields that decouple
from the free scalar fields. The classification problem may then be
focused on subtheories with the restrictive feature that they
decouple from massless scalar free fields. This is the field theoretic
formulation of the result in algebraic QFT \cite{CC05} that every
quantum field theory extension of $\AA$ is contained in $\FF\otimes
\BB$, where $\FF$ is the canonical field net associated with $\AA$ as
above, while $\BB$ is an arbitrary (graded) local net.

The Lie algebras (\ref{ccomm}) and (\ref{rcomm}) are isomorphic to
$\uu(\infty,\infty)$ and $\sp(\infty,\RR)$, respectively. For the
classification of their unitary positive-energy representations, we use methods of highest-weight representations of finite-dimensional
Lie algebras. In fact, we adapt the methods of \cite{EP81,EHW83}
developed for the proof of the Kashiwara--Vergne conjecture
\cite{KV78} (see below) in the finite-dimensional case, and ideas of
\cite{S90} generalizing to the infinite-dimensional case.

Thus, our study relates two independent earlier developments.
The first is the general insight into the structure and origin of
superselection sectors within Haag's operator algebraic approach to
QFT \cite{H92}, culminating in the quoted results of Doplicher and
Roberts \cite{DR90} and subsequent work \cite{CC05}. The second is the
Kashiwara--Vergne conjecture \cite{KV78}, proved by Enright and
Parthasarathy \cite{EP81} and by Jakobsen \cite{J81}, according to which
all unitary highest-weight representations of certain reductive Lie
algebras occur in tensor products of the Segal--Shale--Weil
representation. This amply generalizes the seminal method of
Jordan \cite{J35} and Schwinger \cite{S52} to embed the angular
momentum algebra $\su(2)$ into the algebra of two real harmonic
oscillators. In the present context, it is the analog (for finitely
many degrees of freedom) of our free field Fock representation. 

It should be noted, however, that our assumptions do not precisely
match those of \cite{KV78,EP81,J81}. In the latter, the authors enforce
half-integrality of the Cartan spectrum by demanding integrability 
of the representations of $\uu(n,n)$ and $\sp(2n,\RR)$ to
representations of $\UU(n,n)$ and of the metaplectic group 
$\Mp(2n,\RR)$ (the two-fold covering of $\Sp(2n,\RR)$), respectively.
While we have no direct quantum field theoretic motivation for such 
requirements, it turns out that in our approach the same constraints on 
the spectrum arise either by the choice of the canonical conformal
Hamiltonian, or by the validity of the determinant relations
(\ref{pol}) through the Reeh--Schlieder theorem.

Let us point out that all of our results hold also for
spacetime dimension $D=1$. In this case the bilocal field $W(x_1,x_2)$
generates the vertex algebra $W_{1+\infty}$ with a central charge
$-N$ (corresponding to $+N$ in the terminology of the present article,
Eq.~(\ref{ccomm}), and corresponding to the central charge $c=2N$ as
defined in \cite{NRT05} in terms of the stress-energy tensor).
The representation theory of $W_{1+\infty}$ was developed by Kac and Radul,
and conclusions similar to ours were obtained in \cite{KR96}.

Despite the fact that we do not mention nor use conformal
invariance in the main body of our paper, it should be stressed that
the expansions of the bilocal fields $W(x_1,x_2)$ and $V(x_1,x_2)$
into local fields of twist $D-2$ include the \emph{conformal stress-energy
tensor}. This implies that the conformal Lie algebra $\so(D,2)$
is embedded in the (suitably completed and centrally extended) Lie algebras
$\uu(\infty,\infty)$ and $\sp(\infty,\RR)$, thus generating the global 
conformal invariance of the bilocal fields. In order not to distract
the reader's attention from the main line of argument, we have
relegated the construction of the stress-energy tensor and of the
conformal generators to Appendix~\ref{AB}.

\section{Classification: the complex case}
\label{sec2}\setcounter{equation}{0}

We classify all irreducible unitary positive-energy representations
(superselection sectors) of the Lie algebra (\ref{ccomm}) of the complex
bilocal field $W(x_1,x_2)$.
The completely analogous case of the real bilocal field $V(x_1,x_2)$
will be sketched in the next section. 

\subsection{Statement of the results} 

We first identify the commutation relations (\ref{ccomm}) with the Lie algebra
$\uu(\infty,\infty)$ as follows. We choose an orthonormal basis of the
one-particle space, i.e., a basis of functions $f_i$ $(i=1,2,\ldots)$
of positive energy such that 
\bea \label{fifj}
\Delta_+(\bar f_i,f_j) = \delta_{ij}, \quad
\Delta_+(f_i,\bar f_j) = \Delta_+(f_i,f_j) = \Delta_+(\bar f_i,\bar f_j) = 0.
\eea
Smearing $W(x_1,x_2)$ with these
functions and their conjugates, we define the generators
\bea \label{gen}
&& X_{ij} = W(\bar f_i,\bar f_j), \qquad X_{ij}^* = W(f_j,f_i), \nonumber \\
&& E^+_{ij} = (E^+_{ji})^* = W(f_i,\bar f_j) + \frac
N2\;\delta_{ij},\nonumber \\
&& E^-_{ij} = (E^-_{ji})^* = W(\bar f_j,f_i) + \frac N2\;\delta_{ij} \,,
\qquad i,j=1,2,\ldots \; .
\eea
Then commutation relations (\ref{ccomm}) become equivalent to
the ones of $\uu(\infty,\infty)$ (considered as a Lie algebra over the
complex numbers equipped with the real structure given by the
conjugation properties in (\ref{gen})):
\bea
[E^+_{ij},E^+_{kl}] = \delta_{jk} E^+_{il} - \delta_{il} E^+_{kj}, \qquad
[E^-_{ij},E^-_{kl}] = \delta_{jk} E^-_{il} - \delta_{il} E^-_{kj}, \qquad
[E^+_{ij},E^-_{kl}] = 0, \nonumber
\eea \vskip-9mm
\bea
[E^+_{ij},X^*_{kl}] = \delta_{jl} X^*_{ki},\qquad
[E^+_{ij},X_{kl}] = - \delta_{il} X_{kj},\nonumber \\[1mm]
[E^-_{ij},X^*_{kl}] = \delta_{jk} X^*_{il}, \qquad
[E^-_{ij},X_{kl}] = - \delta_{ik} X_{jl}, \nonumber
\eea \vskip-9mm
\bea \label{ucomm}
[X_{ij},X_{kl}^*] = \delta_{ik} E^+_{lj} + \delta_{jl} E^-_{ki} \;.
\eea
These relations depend neither on the space-time dimension nor on the
parameter $N$ occurring in (\ref{ccomm}), which has been absorbed by
the shift of the generators $E^\pm_{ii}$. The Lie algebra alone
``ignores'' its field theoretic origin (\ref{Wfree}). This observation
could create the impression that the classification problem does not
depend on the parameter $N$ at all. In fact, the parameter $N$ will
reappear either through the additional determinant relations
(\ref{pol}), or through the canonical choice of the Hamiltonian
defining the condition of positive energy, as discussed below.
We say that a representation has \emph{positive energy}
if the Hamiltonian is well defined and diagonalizable, 
its spectrum is bounded from below and all of its eigenspaces are 
finite dimensional. 

\medskip 
{\bf Remark 1:} 
In spite of the fact that the Lie algebra commutation relations
(\ref{ucomm}) are equivalent to the commutation relations
(\ref{ccomm}) of the bilocal field $W(x_1,x_2)$ via (\ref{gen}), it is
not evident if they will fix, up to a unitary equivalence, the field
action. It remains to fix in addition the action of the
\textit{central} modes. Such a situation arises, for instance, in the
case of an abelian current in two-dimensional (chiral) conformal field
theory. In the case at hand of a harmonic bilocal field, the
d'Alembert equation entails that if a mode of the type of (\ref{gen})
is zero in the vacuum representation, then it will be zero also in any
other representation where the d'Alembert equation is satisfied. But
since we are interested only in representations which are locally
intertwined with the vacuum sector, the d'Alembert equation does hold
by virtue of the Reeh-Schlieder theorem. One can easily check the
statement in the mode representation given in Appendix A. Hence, for
harmonic bilocal Lie fields the Lie algebra commutation relations
uniquely fix the whole theory.  
\smallskip

The canonical \emph{conformal} Hamiltonian associated with the free
field expression (\ref{Wfree}) is
\bea \label{Hcan}
H_{\mathrm c}=\sum_{i=1}^\infty \eps_i\cdot (E^+_{ii} + E^-_{ii} - N),
\eea
provided we choose the above orthonormal basis $\{f_i\}$ to
diagonalize the one-particle conformal Hamiltonian with eigenvalues
$\eps_i$. The latter are positive integers and occur with finite
multiplicities depending on the space-time dimension. (An explicit
diagonalization of the conformal Hamiltonian will be provided in
Appendix~\ref{AA}.) 

We also introduce the charge operator 
\bea \label{Q}
Q= \sum_{i=1}^\infty (E^+_{ii} - E^-_{ii}),
\eea
which generates the center of the Lie algebra $\uu(\infty,\infty)$,
and we demand that $Q$ is well defined on the representation.

We define a {\em vacuum representation} as an irreducible unitary
positive-energy representation of the commutation relations
\eqref{ccomm} of the bilocal field $W(x_1,x_2)$, in which 
$Q$ is well-defined and $H_{\rm c}$ has the eigenvalue $0$ on the
ground state $\vac$ (the vacuum state). We shall show in Corollary 1 below
that the vacuum representation exists and is unique, and that
the condition $H_{\rm c}\vac = 0$ is equivalent to the seemingly
stronger requirement that the vacuum expectation value of $W(x_1,
x_2)$ vanishes (and similarly for $V(x_1,x_2)$). 

Now we can state our first main result.

\medskip

\noindent 
{\bf Theorem 1.} {\it Consider irreducible unitary
  positive-energy representations of the commutation relations
  \eqref{ccomm} of the bilocal field\/ $W(x_1,x_2)$, or equivalently
  of the Lie algebra\/ $\uu(\infty,\infty)$, with fixed\/ $N$. We
  assume that the charge operator \eqref{Q} is well defined. The
  condition of ``positive energy'' is with respect to the conformal
  Hamiltonian $H_{\rm c}$ \eqref{Hcan}. Then: 

{\rm{(i)}}
$N$ is a nonnegative integer, and all irreducible unitary positive-energy
representations of\/ $\uu(\infty,\infty)$ are realized (with
multiplicities) in the Fock space of\/ $N$ complex massless free
scalar fields by \eqref{Wfree}.

{\rm{(ii)}} 
The ground states of equivalent representations of\/ $\uu(\infty,\infty)$
in the Fock space form irreducible representations of the gauge group 
$\UU(N)$. This establishes a one-to-one correspondence between the 
irreducible representations of\/ $\uu(\infty,\infty)$ occurring in the 
Fock space and the irreducible representations of\/ $\UU(N)$.
}

\medskip
The case $N=0$ is included, meaning that there is only the
trivial representation.

The important conclusion is that all superselection sectors of the
bilocal field $W(x_1,x_2)$ are realized in the Fock space of $N$
complex massless free scalar fields, as anticipated following the
result of \cite{DR90}, obtained in Haag's framework \cite{H92} using
local nets of von Neumann algebras. Moreover, the multiplicity of a
representation of $W(x_1,x_2)$ in the Fock space equals the dimension
of the corresponding representation of the gauge group $\UU(N)$. We
shall prove Theorem 1 in the remainder of this section. In distinction
to \cite{DR90}, our proof proceeds in a very concrete way based on the
Wightman framework rather than the framework of local von Neumann
algebras. 

A remarkable consequence of Theorem 1 is that the determinant
relations \eqref{detwick} are automatically satisfied in every unitary 
positive-energy representation of the Lie algebra $\uu(\infty,\infty)$.

For completeness, we display the Fock space representation (\ref{Wfree})
of the generators (\ref{gen}) of $\uu(\infty,\infty)$:
\bea \label{fockgen}
X_{ij} = \vec b_i\cdot \vec a_j, \qquad 2E^+_{ij} =
\vec a_i{}^* \cdot\vec a_j + \vec a_j\cdot\vec a_i{}^*, 
\qquad 2E^-_{ij} =
\vec b_i{}^*\cdot\vec b_j + \vec b_j\cdot\vec b_i{}^*\;,
\eea
where $\vec a_i=\vec \varphi(\bar f_i)$ and 
$\vec b_i=\vec\varphi\,{}^{*}(\bar f_i)$
are the annihilation operators for the fields $\vec\varphi$ and
$\vec\varphi{}^{*}$, respectively, and the vector notation indicates
that these fields are multiplets of size $N$. The creation operators are
$\vec a_i{}^* = \vec \varphi\,{}^*(f_i)$ and 
$\vec b_i{}^* = \vec\varphi(f_i)$,
and together with the annihilation operators they satisfy
the canonical commutation relations:
\bea\label{ccr}
\bigl[a_i^p,a_j^q{}^*\bigr] \, = \, 
\delta_{p,q} \delta_{i,j} \, = \, 
\bigl[b_i^p,b_j^q{}^*\bigr],
\quad \bigl[a_i^p,b_j^q{}^{(*)}\bigr] \, = \, 0,\; \text{etc.}
\eea

Next, we give an alternative formulation of the classification problem 
as follows. We start with the assumption that $N$ is a positive integer and 
the bilocal field $W(x_1,x_2)$ has the free field realization (\ref{Wfree})
in the vacuum sector, while all other superselection sectors
are generated from the vacuum one by (relatively) local fields.
We again require that the charge operator (\ref{Q}) be well defined, 
but now we allow a \emph{general} Hamiltonian of 
the form
\bea \label{H}
H=\sum_{i=1}^\infty \eps_i\cdot (E^+_{ii} + E^-_{ii} - g_i).
\eea
It is sufficient to assume that the energies $\eps_i$ form an
increasing sequence of positive numbers,
$0<\eps_1\leq\eps_2\leq\cdots$, with finite degeneracies. The real
parameters $g_i$ replacing the vacuum energy subtractions in
(\ref{Hcan}) may be regarded as a (finite or infinite) additive
renormalization of the Hamiltonian. They will be adjusted in such a
way that the sum (\ref{H}) converges in the representations under
consideration. We shall see in Subsect.\ \ref{detsect} that, using the
Reeh--Schlieder theorem, this proviso eventually fixes the parameters
$g_i$ up to a finite renormalization, which is of course
irrelevant. We thus postulate that an operator (\ref{H}) 
exists and is bounded from below with finite degeneracies.

To apply the Reeh--Schlieder theorem, we consider the operators
\bea \label{detn}
D_n := \det\bigl(X_{ij}\bigr)_{i,j=1}^n \,,
\eea
which arise by smearing the multilocal fields 
$\det (W(x_i,x_j))_{i,j=1}^{n}$ with the conjugates $\bar f_i$
($i=1,\ldots,n$) of the first $n$ basis functions in both arguments.
In the vacuum representation, all determinant operators of the form
(\ref{pol}) vanish, in particular $D_{N+1}$ vanishes, while $D_N\neq 0$. 
(The contraction terms in (\ref{pol}) are absent in this case
because all $\bar f_i$ carry negative energy. Infinitely many other
determinant relations are generated from $D_{N+1}$ by taking
commutators with the generators, but will not be needed for our
argument.) Appealing to the Reeh--Schlieder theorem, we shall require
that $D_{N+1}$ also vanish in the unitary representations of interest,
while $D_N\neq 0$.

\medskip

\noindent 
{\bf Theorem 2.} {\it Consider irreducible unitary positive-energy
representations of the Lie algebra\/ $\uu(\infty,\infty)$
on which the operator\/ $D_{N+1}$ vanishes, while\/ $D_N\neq 0$ 
for some\/ $N\in\NN$. 
Assume that the charge operator \eqref{Q} is well defined, and the
condition of ``positive energy'' holds for the generalized Hamiltonian
\eqref{H}. Then the conformal Hamiltonian \eqref{Hcan} is well defined
and it differs from the generalized one by a finite additive
constant.} 

\medskip

Therefore, the alternative assumptions of Theorem 2 lead to the same
classification as in Theorem 1. The proof of Theorem 1 will
be given in Subsect.\ \ref{cartansect}--\ref{repnsect}, while
Theorem 2 will be proven in Subsect.\ \ref{detsect}.

\subsection{The ground state and the Cartan spectrum} \label{cartansect}

We start with preliminary considerations which hold for a
{\em general} Hamiltonian $H$ of the form \eqref{H}.

Consider an irreducible unitary positive-energy representation 
of the Lie algebra $\uu(\infty,\infty)$ with commutation relations
\eqref{ucomm}. The \emph{Cartan subalgebra} of $\uu(\infty,\infty)$ is
spanned by the generators $E^\pm_{ii}$, which commute with each other
and  with the Hamiltonian $H$. Since, by assumption, $H$
is diagonalizable with finite-dimensional eigenspaces,
it follows that the Cartan generators can be simultaneously diagonalized.  

By the commutation relations, the spectrum of the
Cartan generators $E^\pm_{ii}$ is integer-spaced and in particular discrete.
A joint eigenvalue is a pair of sequences
$\ul h^+ = (h^+_1,h^+_2,\ldots)$, $\ul h^-=(h^-_1,h^-_2,\ldots)$. We
denote such a pair by $\ul h = (\ul h^+,\ul h^-)$. On a state $\ket h$
with eigenvalues $h^\pm_i$ of $E^\pm_{ii}$, 
the Hamiltonian $H$ \eqref{H} has
the eigenvalue $\sum_i\eps_i(h^+_i+h^-_i-g_i)$.

Since $X_{ij}$ lowers the eigenvalues of $H$ by
$\eps_i+\eps_j>0$ and $H$ is bounded from below, there must be a
\emph{ground state} $\ket h$ annihilated by all $X_{ij}$.
If $\eps_i<\eps_j$, then $E^\pm_{ij}$ lowers the eigenvalues of
$H$ by $\eps_i-\eps_j$, hence these elements
also annihilate the ground state. If
$\eps_i=\eps_j$, then $E^\pm_{ij}\ket h$ has the same eigenvalue as
$\ket h$, i.e., the ground state may be degenerate. Let an eigenvalue
$\eps$ of the one-particle Hamiltonian be $n$-fold degenerate
($n<\infty$).
Then the ground states form a representation of the Lie subalgebra
$\uu(n)\oplus \uu(n)$ with generators $E^\pm_{ij}$
($\eps_i=\eps_j=\eps$). 
Choose for $\ket h$ a highest-weight vector of this representation, so
that $E^\pm_{ij}\ket h=0$ whenever $i<j$
($\eps_i=\eps_j=\eps$). Because $E^\pm_{ij}$ and $E^\pm_{kl}$ 
commute whenever $\eps_i=\eps_j\neq\eps_k=\eps_l$, the same can be
done for all degenerate eigenvalues of the one-particle Hamiltonian
simultaneously. 

Thus the ground state $\ket h$ can be chosen to satisfy
\bea \label{hw}
X_{ij}\ket h=0\quad\forall\; i,j,\qquad E^\pm_{ij}\ket h=0
\quad\hbox{for}\; i<j, \qquad \hbox{and}\qquad E^\pm_{ii}\ket h =
h^\pm_i\;\ket h.
\eea

Together with the commutation relations, the pair of sequences
$\ul h = (\ul h^+,\ul h^-)$ of Cartan eigenvalues determines the inner
product and hence the representation completely. Unitarity imposes
conditions on $\ul h$, some of which are elementary to obtain. 
Computing 
\bea \label{recX}
\bra h X_{ij}X_{ij}^* \ket h = h^+_j+h^-_i
\eea  
we deduce from unitarity that $h^+_j + h^-_i$ must be nonnegative for 
all $i,j$. Computing 
\bea \label{recE1}
\bra h E^\pm_{ij}E^\pm_{ji}\ket h =h^\pm_i-h^\pm_j \,,
\qquad i<j, 
\eea
we conclude that both sequences $\ul h^\pm =
(h^\pm_1,h^\pm_2, \ldots)$ are weakly decreasing. Computing further
recursively 
\bea \label{recE}
\bra h (E^\pm_{ij})^n(E^\pm_{ji})^n \ket h =n! \;
(h^\pm_i-h^\pm_j)(h^\pm_i-h^\pm_j-1)\cdots(h^\pm_i-h^\pm_j-n+1),
\eea
we obtain from unitarity that all differences $h^\pm_i-h^\pm_j$ are
nonnegative integers and 
\bea \label{null}
(E^\pm_{ji})^{h^\pm_i-h^\pm_j+1}\ket h=0, \qquad i<j.
\eea
Thus, the eigenvalues $\ul h^\pm$ form a pair of integer-spaced
weakly decreasing sequences such that $h^+_i+h^-_i\geq 0$. 
Therefore, both sequences $\ul h^+$ and $\ul h^-$ must stabilize at 
some values $h^+_\infty$, $h^-_\infty$. The convergence of the
eigenvalue $\sum_i (h^+_i-h^-_i)$ of $Q$ on the ground state implies
that 
\bea \label{hpminfty}
h^+_\infty = h^-_\infty =:h_\infty \geq 0.  
\eea

If we now specialize to the {\em canonical} Hamiltonian
(\ref{Hcan}), then the convergence of the eigenvalue $\sum_i
\eps_i\cdot(h^+_i+h^-_i-N)$ of $H_{\rm c}$ implies 
\bea\label{hpminftyN}
2h_\infty=N.
\eea

To prove that $N$ is a nonnegative integer, let $r$ be
sufficiently large such that $h^+_i=h^-_i=h_\infty$ for $i>r$, and
observe that by Eqs.~\eqref{hw}, \eqref{recE1} one has
$E_{ij}^{\pm}\,\ket h = h_\infty \;\delta_{ij}\;\ket h$ for all 
$i,j > r$.   
This allows one to compute recursively the norm square of $D^{(r)}_n{}^*\ket
h$, where $D^{(r)}_n:= \det\bigl(X_{ij}\bigr)_{i,j=r+1}^{r+n}$. 
Instead of computing it explicitly, it suffices to observe as in
\cite{NST02} that this norm square must be a polynomial $p_n(N)$ of
degree $n$ in $N=2h_\infty$, and that $p_n(N)$ vanishes whenever $N$ is
a nonnegative integer smaller than $n$. Indeed, if $N\in\NN_0$, the
representation under consideration, restricted to the subalgebra with
generators $X_{ij}$, $X^*_{ij}$ and $E^\pm_{ij}$ with $i,j>r$, 
coincides with the vacuum representation of the free-field realization
\eqref{Wfree} with $N$ complex scalar fields, where $D^{(r)}_n$ vanishes
manifestly for $n>N$. For the same reason, for $n=N$, $p_n(n)$ is
positive. These facts taken together imply that 
\bea 
p_n(N) \sim N(N-1)\cdots(N-n+1) 
\eea 
with an irrelevant positive coefficient. Then it is clear that nonnegativity 
of all $p_n(N)$ for a given value of $N=2h_\infty$ implies that 
$N$ must be a nonnegative integer. 
%The case $N=0$ is excluded since in this case the vacuum
%representation is identically zero.  Therefore, $N\in\NN$.

In particular, if we define the vacuum state requiring $H_{\rm c}\vac =0$, 
we must have $h^\pm_i=h_\infty =  N/2$ for all $i$. Because the
sequence of Cartan weights determines the representation, the vacuum
representation is unique, and given by Eq.~\eqref{fockgen}. Moreover,
we have the following result. 

\medskip

\noindent 
{\bf Corollary 1.} {\it The vacuum state\/ $\vac$ spans a
  one-dimen\-sional representation of the Lie subalgebra\/
  $\uu(\infty)\oplus\uu(\infty)$ with generators\/ $E^\pm_{ij}$, such
  that   
\bea \label{vacuum}
E_{ij}^{\pm}\,\vac = \frac N2 \;\delta_{ij}\;\vac.
\eea 
In particular, $Q=0$, and the vacuum expectation value of the
bilocal field vanishes,  
\bea\label{vacW}
\cav W(x_1,x_2)\vac =0. 
\eea
}

Note that \eqref{vacW} for a ground state $\vac$ implies 
$\cav H_{\rm c} \vac=0$ and hence $H_{\rm c}\vac = 0$ (see Proposition
2 in Appendix \ref{AB}). One could have expected \eqref{vacW} to be a
part of the definition  of the bilocal field as described in the
introduction, but we see here that it follows from the seemingly
weaker assumption that the vacuum state has zero energy. 

\subsection{Unitarity bounds from Casimir operators}
\label{casisect}

To obtain further constraints on the admissible values of $\ul h$, we
shall consider certain finite-dimensional subalgebras of $\uu(\infty,\infty)$.
Namely, for a positive integer $n$, we consider the Lie subalgebra $\uu(n,n)$
spanned by the generators (\ref{gen}) with indices $1\leq i,j\leq n$. 
We choose $n$ sufficiently large so that $h^+_n=h^-_n=h_\infty$.

Clearly, unitarity of a representation of $\uu(\infty,\infty)$ implies
unitarity of its restriction to $\gg:=\uu(n,n)$. We may then follow the
strategy of \cite{EP81}, using results of \cite{EHW83}.
Denote by $\kk:= \uu(n) \oplus \uu(n)$
the Lie algebra of the maximal compact subgroup
$\UU(n)\times \UU(n)$ of $\UU(n,n)$. 

We adapt the conventions of
\cite{EP81} for the positive roots of these Lie algebras so that our
\emph{lowest energy} condition $h^\pm_1\geq\cdots\geq h^\pm_n$ turns into
a \emph{highest weight} condition. 
Introduce an orthonormal basis $\ul e^\pm_i$
($i=1,\ldots,n$) of $\RR^{2n}$ such that $\ul h = \sum_i(h^+_i\ul
e^+_i + h^-_i\ul e^-_i )$. 
We define the positive roots to be
the roots $\ul e^\pm_i- \ul e^\pm_j$ ($i<j$)
and $-\ul e^-_i-\ul e^+_j$ associated with the annihilation operators
$E^\pm_{ij}$ ($i<j$) and $X_{ij}$ for the ground state; 
then the ground state $\ket h$ is a highest-weight vector for $\gg$.

Unitarity of an irreducible representation $U_\gg(\ul h)$ 
of $\gg$ with a ground state $\ket h$ 
is equivalent to the condition that the inner product on the Verma
$\gg$-module $V_\gg(\ul h)$ is semi-definite. 
Then $U_\gg(\ul h) = V_\gg(\ul h)/N_\gg(\ul h)$ 
is the quotient of the Verma module by its (maximal) submodule of null
vectors.

We also introduce the Verma $\kk$-module $V_\kk(\ul h)$ and its
quotient $U_\kk(\ul h) = V_\kk(\ul h)/N_\kk(\ul h)$ by the maximal
submodule of null vectors $N_\kk(\ul h) = N_\gg(\ul h) \cap V_\kk(\ul h)$.
In fact, $N_\kk(\ul h)$ is generated by the null vectors from
Eq.~(\ref{null}), and $U_\kk(\ul h)$ is the unitary representation of
$\kk$ specified as follows.
Denote by $\ul h^\pm$ the finite sequence
$(h^\pm_1,\ldots,h^\pm_n)$. Then we have 
\bea \label{m}
h^\pm_i- h_\infty =m^\pm_i
\eea
where 
\bea \label{Y}
m^\pm_1\geq m^\pm_2 \geq \cdots \geq m^\pm_{r^\pm}>
m^\pm_{r^\pm+1}=\cdots = m_n = 0\quad\hbox{are
  nonnegative integers}.
\eea
The restriction of the representation $U_\kk(\ul h)$ to $\su(n)\oplus
\su(n)\subset \kk$ is given by the pair of Young diagrams $Y^\pm$
with $r^\pm$
rows of length $m^+_i$ and $m^-_i$ ($1\leq i\leq r^\pm < n$), respectively,
while $h_\infty$ determines the representation of the center
$\uu(1)\oplus \uu(1)$ of $\kk$. 

To get a necessary condition for the unitarity of $U_\gg(\ul h)$,
we exploit the eigenvalues of \emph{Casimir operators}.
The Casimir operator for $\kk$, 
\bea \label{Ck1}
C_\kk=\sum\nolimits_{ij}(E^+_{ij}E^+_{ji} + E^-_{ij}E^-_{ji}), 
\eea
has an eigenvalue $(\ul\lambda +
\ul\varrho,\ul\lambda + \ul\varrho)-(\ul\varrho,\ul\varrho)$ in any
highest-weight representation of $\kk$ with highest weight $\ul\lambda$, where
$\ul\varrho=\frac12\sum_{i=1}^n(n+1-2i)(\ul e^+_i+\ul e^-_i)$ is one-half
the sum of all positive roots of $\kk$. $(\cdot,\cdot)$ is the natural
inner product in $\RR^{2n}$.

On the other hand, the Casimir operator for $\gg$,
\bea \label{Cg1}
C_\gg=\sum\nolimits_{ij}(E^+_{ij}E^+_{ji} + E^-_{ij}E^-_{ji} - X_{ij}^*X_{ij} -
X_{ij}X_{ij}^*),
\eea 
has an eigenvalue $(\ul h + \ul\delta,\ul h +
\ul\delta)-(\ul\lambda,\ul\lambda)$ in any highest-weight
representations of $\gg$ with highest weight $\ul h$, where $\ul\delta =
\ul\varrho-\frac n2\sum_{i=1}^n(\ul e^+_i+\ul e^-_i)$.

Then writing the difference as $C_\kk- C_\gg = 2\sum_{ij}
X_{ij}^*X_{ij}+\sum_i(E^+_{ii}+E^-_{ii})$, it is easy to calculate
\bea \label{casimir}
2\sum\nolimits_{ij}\bra\lambda X_{ij}^*X_{ij}\ket\lambda =
\big[(\ul\lambda+\ul\delta,\ul\lambda+\ul\delta)-(\ul h+\ul\delta,\ul
h+\ul\delta) \big]\cdot\braket\lambda\lambda ,
\eea
whenever $\ket\lambda$ is a highest-weight vector for $\kk$ of weight
$\ul\lambda$ within a highest-weight $\gg$-module with highest
weight $\ul h$.

Assume now that the highest weight $\ul h$ gives rise to an
irreducible unitary representation of $\gg$. Then $U_\kk(\ul h)$
induces the highest-weight $\gg$-module $M(\ul h) =
\CC[X_{kl}^*]\otimes U_\kk(\ul h)$ spanned by vectors of the
form $X^*\cdots X^* E^\pm\cdots E^\pm \ket h$, 
which is the quotient of the Verma
module $V_\gg(\ul h)$ by the (nonmaximal in general) submodule
$\CC[X_{kl}]\otimes N_\kk(\ul h)$.

Consider the subspace $\CC \{ X_{kl}^* \} \otimes U_\kk(\ul h)$ of $M(\ul h)$
spanned by vectors of the form $X^* E^\pm\cdots$ $E^\pm\ket h$. This is a
$\kk$-module equivalent to $U_\kk(\ul e^+_1 + \ul e^-_1) \otimes
U_\kk(\ul h)$, because the generators $X_{kl}^*$ transform like a vector 
with respect to both
$\su(n)$ factors of $\kk$. Let $\ul\lambda$ be the highest
weight of any $\kk$-subrepresentation of $U_\kk(\ul e^+_1 + \ul e^-_1)
\otimes U_\kk(\ul h)$, and let $\ket\lambda$ be the corresponding vector in
$\CC \{ X_{kl}^* \} \otimes U_\kk(\ul h)\subset M(\ul h)$.

By the unitarity assumption, expression (\ref{casimir})
must be nonnegative, and if it is positive, then
$\braket\lambda\lambda >0$ and
$(\ul\lambda+\ul\delta,\ul\lambda+\ul\delta) > (\ul h+\ul\delta,\ul
h+\ul\delta)$. If instead {(\ref{casimir})} vanishes, then
$\bra\lambda X_{ij}^*X_{ij}\ket\lambda$ must vanish for all $i,j$.
Since in $M(\ul h)$ we have 
$X_{ij} U_\kk(\ul h) = \{0\}$, the commutation relations \eqref{ucomm}
imply that the generators $X_{ij}$ map any element of 
$\CC \{ X_{kl}^* \} \otimes U_\kk(\ul h)$
into $U_\kk(\ul h)$. In particular, $X_{ij}\ket\lambda$ belongs to the
Hilbert space $U_\kk(\ul h)$. Hence $X_{ij}\ket\lambda=0$, 
and therefore $\ket\lambda$ is a highest-weight vector for $\gg$
with weight $\ul \lambda$ within $M(\ul h)$. This implies that
$\braket\lambda\lambda=0$ and that $\ul\lambda + \ul\delta$ is
a $\gg$-Weyl transform of $\ul h + \ul\delta$ \cite{V68,BGG71}. 
Then 
$(\ul\lambda+\ul\delta,\ul\lambda+\ul\delta) = (\ul h+\ul\delta,\ul
h+\ul\delta)$. We have obtained in both cases
\bea \label{gamma}
(\ul\lambda+\ul\delta,\ul\lambda+\ul\delta)-(\ul
h+\ul\delta,\ul h+\ul\delta) =:\gamma \geq 0.
\eea

By the Littlewood--Richardson rule, $U_\kk(\ul e^+_1 + \ul e^-_1)
\otimes U_\kk(\ul h)$ contains $U_\kk(\ul\lambda)$ with $\ul\lambda^\pm =
\ul h^\pm+\ul e_{r^\pm+1}$ where $r^\pm$ are the heights of
the first columns of the Young diagrams $Y^\pm$ defined by $\ul h^\pm$
according to (\ref{Y}). 
For this choice of $\ul\lambda$ we have $\gamma = 2(2h_\infty -r^+-r^-)$
(it can be shown that this is the minimal value of $\gamma$).
Then (\ref{gamma}) gives the following necessary condition for unitarity:
\bea \label{bound}
r^++r^-\leq 2h_\infty.
\eea

\subsection{Fock space representations}
\label{focksect}

Kashiwara and Vergne \cite{KV78} have shown that all highest-weight
representations of $\su(n,n)$ satisfying the bound (\ref{bound}) with
$h^+_n=h^-_n = h_\infty$ half-integer are contained in the $(r^++r^-)$-fold
tensor power of the Segal--Shale--Weil representation, and Schmidt
\cite{S90} has extended this result to $n=\infty$. We essentially
reformulate these results in our setting.

Recall that the bilocal field $W(x_1,x_2)$ is given on the Fock space by
(\ref{Wfree}). Consequently, the generators
(\ref{gen}) of $\uu(\infty,\infty)$ are given by (\ref{fockgen}). Clearly,
the representation of $\uu(\infty,\infty)$ on the Fock space is unitary.

We claim that every representation with a ground state $\ket h$
satisfying the bound \eqref{bound} with $2h_\infty = N\in\NN$ 
is contained in the Fock space of $N$ complex free scalar fields. 
(The case $N=0$ implies $r^+=r^-=0$, and hence triviality of the
representation by virtue of Eqs.~\eqref{recX} and \eqref{recE1}.)

It is sufficient to display a vector with the properties of
$\ket h$ within the Fock space. Let $h^\pm_i = m^\pm_i+N/2$
according to (\ref{m}) (with $n$ sufficiently large), and let $Y^\pm$ be the
associated Young diagrams with rows of length $m^\pm_i$. Denote
the heights of the columns of these diagrams by
$r^\pm = r^\pm_1\geq r^\pm_2\geq\cdots\geq r^\pm_{m^\pm_1}$.

Consider the Fock space vector 
\bea \label{fockground}
\ket h_F = \Biggl( \prod_{k=1}^{m^+_1}
a^{*\wedge r^+_k} \Biggr) \Biggl( \prod_{l=1}^{m^-_1}
b^{*\wedge r^-_l} \Biggr) \vac
\eea
where $a^{*\wedge r}$ and $b^{*\wedge r}$ stand for the components 
\bea a^{*\wedge r} = \det \Big(a_i^p{}^*\Big){}_{p=1,\ldots, r \atop
  i=1,\ldots, r} \qquad\hbox{and}\qquad b^{*\wedge r} = 
\det \Big(b_i^p{}^*\Big){}_{p=N+1-r,\ldots, N \atop i=1,\ldots, r\;\qquad}
\eea
of the antisymmetric $\UU(N)$ tensors $\vec c_1{}^*\wedge\cdots
\wedge \vec c_r{}^*$ ($c=a$ or $b$). Note that $\vec a_i{}^*$ and
$\vec b_i{}^*$ transform like a $\UU(N)$ vector and a conjugate
vector, respectively (see Subsect.\ \ref{repnsect}). The vector
(\ref{fockground}) is annihilated by $X_{ij}$ for all $i,j$ (because 
$r^+_k\leq N-r^-_l$ thanks to the bound
(\ref{bound}) and $2h_\infty=N$) and by $E^\pm_{ij}$ for $i<j$ (by virtue 
of the antisymmetrizations). In addition, $\ket h_F$
has eigenvalues $h^\pm_i$ for the 
operators $E^\pm_{ii}$. Therefore, $\ket h_F$ has all the properties of the 
ground state $\ket h$. This proves 
part (i) of  Theorem 1. 

Because the Fock space representation is unitary, the presence of this
ground state in the Fock space proves, in particular, that the
necessary conditions for unitarity found above are in fact also
sufficient \cite{KV78}. 

We conclude that all superselection sectors of the complex bilocal field
$W(x_1,x_2)$ are realized in the Fock space of $N$ complex massless
scalar fields. The sectors are classified by
the Cartan eigenvalues of $\uu(\infty,\infty)$:
\begin{equation}\label{data}
\begin{split}
\ul h^+ &= (m^+_1+h_\infty,\ldots, m^+_{r^+}+h_\infty, h_\infty,\ldots),
\\
\ul h^- &= (m^-_1+h_\infty,\ldots, m^-_{r^-}+h_\infty, h_\infty,\ldots),
\end{split}
\end{equation}
where $h_\infty =N/2$, $m^\pm_1\geq\cdots\geq m^\pm_{r^\pm}>0$ are
integers, and $r^++r^-\leq N$. 

\subsection{Representations of the gauge group}
\label{repnsect}

It remains to relate the superselection sectors of the bilocal field
$W(x_1,x_2)$ classified in the previous subsections to the
unitary representations of the gauge group $\UU(N)$. 

The gauge group $\UU(N)$ is unitarily represented on the Fock space in
such a way that the vacuum is invariant and the creation operators
$\vec a\,{}^*$ and $\vec b\,{}^*$ transform like an $N$-vector and a
conjugate $N$-vector, respectively. In particular, the expressions
(\ref{fockgen}) and the bilocal fields $W(x_1,x_2)$ given by (\ref{Wfree})
on the Fock space are gauge invariant. Because the gauge group and the
fields commute with each other, it follows that a ground state
$\ket h_F$ is a component of a $\UU(N)$ tensor representation whose
dimension equals the multiplicity of the corresponding superselection
sector within the Fock space.  

The ground state $\ket h_F$ displayed in (\ref{fockground}) is in fact
a common highest weight vector for the commuting actions of
$\uu(\infty,\infty)$ and $\uu(N)$ on the Fock space. The latter is
the Lie algebra of the gauge group with generators 
\bea \label{eab}
2E^{pq} = \sum_{i=1}^\infty (a_i^p{}^*\, a_i^q + a_i^q \,a_i^p{}^*
- b_i^q{}^*\, b_i^p - b_i^p\, b_i^q{}^*)
\eea
which annihilate $\ket h_F$ if $p<q$. (The infinite sum in (\ref{eab})
converges on all Fock vectors on which $Q$ is finite.)

The components $\prod_k a^{*\wedge r^+_k}$ and 
$\prod_l b^{*\wedge r^-_l}$ in (\ref{fockground})
belong to tensors transforming under the
gauge subgroup $\SU(N)$ in the representations given by the Young
diagrams $Y^+$ and $(Y^-)^*$, respectively. (The latter is the
conjugate diagram whose columns have heights $N-r^-_{m_1^-}\geq \cdots
\geq N-r^-_1$.) The ground state $\ket h_F$ therefore belongs to a
subrepresentation of the tensor product $Y^+ \otimes (Y^-)^*$. The
tracelessness of (\ref{fockground}) implies that the only contribution
comes from the irreducible representation whose Young diagram $Y$ has
column heights $N-r^-_{m_1^-}\geq \cdots \geq N-r^-_1 \geq r^+_1 \geq
\cdots \geq r^+_{m^+_1}$, obtained as the juxtaposition of $(Y^-)^*$ and
$Y^+$ (recall that $N-r^-_1\geq r^+_1$ by (\ref{bound}) 
and $2h_\infty=N$).

Similarly, because $\vec a\,{}^*$ carries
charge $1$ and $\vec b\,{}^*$ carries charge $-1$, the $\UU(1)$
transformation is specified by the eigenvalue of the charge operator
$Q$ on $\ket h_F$ given by $q=\vert Y^+\vert - \vert Y^-\vert = \vert
Y\vert - N\,m^-_1 $, where $\vert Y\vert$ stands for the number of
boxes of the diagram $Y$ so that $\vert (Y^-)^*\vert = N\,m_1^- -\vert
Y^-\vert$. Therefore, the $\UU(N)$ transformation given by the pair
$(Y,q)$ is determined by the pair $(Y^+,Y^-)$.

Conversely, every irreducible unitary representation of $\UU(N)$ is
given by a pair $(Y,q)$ where $\vert q\vert \leq\vert Y\vert$ and $q =
\vert Y\vert \mod N$. (Each $\UU(N)$ vector contributes $1$ to $\vert Y
\vert$ and $1$ to $q$, while each conjugate vector contributes $N-1$
to $\vert Y \vert$ and $-1$ to $q$.) The pair $(Y,q)$ then determines
a unique split of $Y$ into $Y^+$ and $(Y^-)^*$ such that $q=\vert
Y^+\vert - \vert Y^-\vert$.

This gives an explicit one-to-one correspondence between the data
(\ref{data}) defining the superselection sectors and the unitary
irreducible representations of the gauge group $\UU(N)$. Moreover,
since the Cartan eigenvalues determine the occupation numbers in the
ground state, one can see that the above ground states
(\ref{fockground}) exhaust the multiplicity of the superselection
sector representation in the Fock space, and hence the multiplicity space
carries the corresponding representation of $\UU(N)$. This proves 
part (ii) of Theorem 1.

\subsection{The determinant relations}
\label{detsect} 

To prove Theorem 2, we follow the line of argument of Subsect.\ 
\ref{cartansect} until \eqref{hpminfty} for which we did not need the special 
choice $H_{\rm c}$ of the Hamiltonian. 

We now assume that for some $N\in\NN$ the operator $D_{N+1}$ vanishes
in the representation under consideration, while $D_n$ ($n\leq N$) do
not vanish. It is easy to compute, using expansion formulas for
determinants and the commutation relations (\ref{ucomm}), that
\bea \label{D}
X_{nn}D_{n}^*\ket h = (h^+_n+h^-_n-n+1)D_{n-1}^*\ket h .
\eea
Hence
\bea \label{recD}
X_{11}\cdots X_{nn}D_{n}^*\ket h =
\Big(\prod_{m=1}^{n}(h^+_m+h^-_m-m+1)\Big)\ket h ,
\eea
and $D_{N+1}$ can vanish only if
\bea \label{nu}
h^+_m+h^-_m = m-1 \quad\hbox{for some positive integer}\quad m\leq N+1.
\eea
In particular, since the Cartan eigenvalues are integer-spaced and
$h^+_\infty=h^-_\infty$, we conclude that all Cartan eigenvalues
belong to $\frac12\NN_0$. Setting $N':=2h_\infty$, we obtain $N' \leq
h^+_m+h^-_m=m-1\leq N$.

Since the representation under consideration is determined by its
Cartan eigenvalues $\ul h^\pm$, and $h_\infty = N'/2$, we know from
Subsect.\ \ref{casisect} and \ref{focksect} that it is realized on the
Fock space of $N'$ complex scalar fields. But in the Fock
representation, $D_n$ vanishes if $n>N'$ while $D_n$ ($n\leq N'$) are
nontrivial. Therefore our assumption that $D_N$ does not vanish
implies that $N\leq N'$. We conclude that $N'=N$ and 
\bea \label{hinfty}
2h_\infty=N\;.
\eea

Turning to the Hamiltonian \eqref{H}, we observe that its eigenvalue 
on the ground state can converge only if $\sum_i\eps_i(2h_\infty-g_i)$ is 
finite. Hence we must have 
\bea \label{g}
g_i=2h_\infty + \delta_i = N+\delta_i \,
\eea
with arbitrary shifts $\delta_i$ such that $\Delta E = \sum_i\eps_i\delta_i$ 
is well-defined. Clearly, this constitutes just an irrelevant additive 
renormalization $H = H_{\rm c} -\Delta E$ of the Hamiltonian. 

This proves Theorem 2.

\section{Classification: the real case}
\label{sec3}\setcounter{equation}{0}

The classification of the superselection sectors of the real bilocal field
$V(x_1,x_2)$ satisfying (\ref{rcomm}) 
proceeds in perfect analogy to the complex case discussed in the previous
section. We shall just repeat the relevant steps and point out the differences.

The generators of the Lie algebra are
\bea
X_{ij} = V(\bar f_i,\bar f_j) = X_{ji} , \qquad X_{ij}^*\qquad\hbox{and}\qquad
E_{ij} = E_{ji}^* = V(f_i,\bar f_j) + \frac N2\;\delta_{ij}.
\eea
They satisfy the commutator relations of $\sp(\infty,\RR)$:
\bea
[E_{ij},E_{kl}] &=& \delta_{jk} E_{il} - \delta_{il} E_{kj}, \nonumber
\\[1mm] 
[E_{ij},X^*_{kl}] = \delta_{jk} X^*_{il} + \delta_{jl} X^*_{ki}, &&
[E_{ij},X_{kl}] =  - \delta_{ik} X_{jl}- \delta_{il} X_{kj}, \nonumber
\\[1mm] 
[X_{ij},X_{kl}^*] &=& \delta_{jk} E_{li} + \delta_{jl} E_{ki} +
\delta_{ik} E_{lj} + \delta_{il} E_{kj} \;.
\eea
The Fock space representation is given by
\bea \label{fockgenreal}
X_{ij} = \vec a_i\cdot\vec a_j \qquad\hbox{and}\qquad 2E_{ij}
= \vec a_i{}^* \cdot\vec a_j + \vec a_j\cdot\vec a_i{}^* \;,
\eea
where $\vec a_i=\vec \varphi(\bar f_i)$ and 
$\vec a_i{}^* = \vec \varphi(f_i)$
(cf.\ \eqref{ccr}).

The \emph{general} Hamiltonian is
\bea \label{rH}
H=\sum_{i=1}^\infty \eps_i\cdot (E_{ii} - g_i),
\eea
while the canonical \emph{conformal} Hamiltonian is
\bea \label{rHc}
H_{\mathrm c}=\sum_{i=1}^\infty \eps_i\cdot \Bigl( E_{ii} - \frac{N}2 \Bigr).
\eea
There is no charge operator in the real case. 
The determinant operators $D_n$ are defined by
the same formula (\ref{detn}) as in the complex case.

\medskip

\noindent
{\bf Theorem 3.} {\it 
The statements of Theorems 1 and 2 hold if we replace everywhere the
bilocal field\/ $W(x_1,x_2)$ by $V(x_1,x_2)$, the Lie algebra
$\uu(\infty,\infty)$ by $\sp(\infty,\RR)$, complex free fields by real
free fields, the gauge group $\UU(N)$ by $\OO(N)$, and omit the
assumption about the charge operator.} 

\medskip

The important conclusion is that all superselection sectors
are realized in the Fock space of $N$ real massless free scalar fields 
by (\ref{Vfree}).
In the remainder of this section we give a sketch of the proof
of the theorem.

The ground state $\ket h$ is annihilated by all $X_{ij}$ and by
$E_{ij}$ for $i<j$. Computing the same norms as in Subsect.\ \ref{cartansect}, 
we conclude that the Cartan eigenvalues of the generators $E_{ii}$
are given by a single integer-spaced sequence $\ul h$ such that 
$h_1\geq h_2\geq\cdots\geq 0$. This sequence must stabilize
at some value $h_\infty$. Finiteness of the canonical Hamiltonian
$H_{\mathrm c}$ requires $2h_\infty=N$. 
Exploiting the vanishing of $E_{ij}$ on the ground state whenever
$h_i=h_j= h_\infty$ and $i\neq j$, we can determine the norms of the vectors
$\det\bigl(X_{ij}\bigr)_{i,j=r+1}^{r+n} \;\ket h$ and conclude that if
they are nonnegative, then $N$ must be a nonnegative integer. The
unique vacuum representation is given by \eqref{fockgenreal}, and the
obvious analog of Corollary 1 holds.

For a positive integer $n$ such that $h_n=h_\infty$, consider the
restriction of our representation of $\sp(\infty,\RR)$ to a
unitary representation of the
maximal compact subalgebra $\kk:= \uu(n)$ of $\gg:=\sp(2n,\RR)\subset
\sp(\infty,\RR)$.
Then the Cartan eigenvalues have the form $h_i=m_i+h_\infty$
($i\leq n$), where $m_1\geq\cdots\geq m_r>0$ are integers and
$m_{r+1}=\cdots=m_n=0$. The Young diagram $Y$ with rows of lengths
$m_i$ determines the representation of $\su(n) \subset \uu(n)$ 
with highest weight
$\sum m_i\ul e_i$, while $h_\infty$ determines the action of the center of
$\uu(n)$. Considering the Casimir operators 
\bea \label{Ck2}
C_\kk=\sum\nolimits_{ij}E_{ij}E_{ji}
\eea
of $\kk$ and 
\bea \label{Cg2}
C=\sum\nolimits_{ij} \Bigl( E_{ij}E_{ji} -
\frac12(X_{ij}^*X_{ij}+X_{ij}X_{ij}^*) \Bigr)
\eea
of $\gg$, one arrives at
\bea
\sum\nolimits_{ij}\bra\lambda X_{ij}^*X_{ij}\ket\lambda = \gamma\cdot
\braket\lambda\lambda , \qquad
\gamma = (\ul\lambda+\ul\delta,\ul\lambda+\ul\delta)-(\ul
h+\ul\delta,\ul h+\ul\delta),
\eea
whenever $\ket\lambda$ is a highest-weight vector for $\kk$ of weight
$\ul\lambda$ within a highest-weight $\gg$-module with highest
weight $\ul h$. Here, $\ul\delta = -\sum_{i=1}^n i\cdot\ul e_i$. By the 
same argument as in Subsect.\ \ref{casisect}, $\gamma$ is nonnegative.

The adjoint representation of $\kk$ on the linear span of $\{ X_{kl}^* \}$ 
is given by $U(2\ul
e_1)$, hence we may choose for $\ul\lambda$ the highest weight of
any irreducible subrepresentation of the tensor product
$U(2\ul e_1)\otimes U(\ul h)$. By the Littlewood--Richardson rule, we
may choose $\ul\lambda = \ul h + \ul e_{r+1} + \ul e_{s+1}$, where 
$r$ and $s\leq r$ are the heights of the first two columns of the Young
diagram $Y$ (i.e., $r$ is the smallest number such
that $h_{r+1}=h_\infty$ and $s$ is the smallest number such that
$h_{s+1}\leq h_\infty+1$). This choice of $\ul\lambda$ 
gives the necessary condition for unitarity 
\bea\label{rnsn}
r+s\leq 2h_\infty = N.
\eea

Next, we display a ground
state $\ket h_F$ in the Fock space of $N$ real scalar fields by 
\bea
\ket h_F = \Biggl[\prod_{k=1}^{m_1} a^{*\wedge r_k} \Biggr]^0 \vac,
\eea
where $r_k$ are the heights of the columns of the Young diagram $Y$ 
and $a^{*\wedge r}$ stands for the component 
\bea a^{*\wedge r} = 
\det \Big(a_i^p{}^*\Big){}_{p=1,\ldots, r \atop i=1,\ldots, r} 
\eea
of the antisymmetric $\OO(N)$ tensor $\vec a_1{}^*\wedge\cdots\wedge 
\vec a_r{}^*$ ($r\leq N$), while $[\cdots]^0$ stands for the 
corresponding component of the traceless part of the product tensor. 

The presence of this ground state implies that all representations
with highest weights as specified above are realized in this Fock
space, and are indeed unitary. The superselection sectors of the
bilocal field $V(x_1,x_2)$  are thus classified by the Cartan
eigenvalues of $\sp(\infty,\RR)$:
\bea \label{odata}
\ket h = (m_1+h_\infty,\ldots, m_r+h_\infty,h_\infty,\ldots)
\qquad\hbox{with}\quad r+s\leq N,
\eea
where $h_\infty=N/2$, $m_1\geq m_2\geq\cdots\geq m_r>0$ are
integers, and $r$ and $s$ are the heights of the first two columns of
the Young diagram $Y$ whose rows have lengths $m_i$.

The gauge group $\OO(N)$ acts unitarily on the Fock space by
leaving the vacuum invariant and transforming the creation operators 
$\vec a\,{}^*$ like a vector. Therefore the ground state $\ket h_F$
belongs to the unitary representation of $\OO(N)$ given by the Young
diagram $Y$. In fact, it is a common highest-weight vector 
for the commuting actions of $\sp(\infty,\RR)$ and $\so(N)$ (the Lie
algebra of the gauge group) on the Fock space.

By the unitarity bound (\ref{rnsn}), only those Young diagrams occur
whose first two columns have total height $r+s\leq N$. 
It remains to convince oneself that such Young diagrams give precisely
all irreducible unitary representations of $\OO(N)$. The standard
labeling \cite{B70} of the unitary representations of $\OO(N)$ is
given by pairs $(Y,\pm)$ where $Y$ is a Young diagram with at most
$N/2$ rows determining the representation of the subgroup
$\SO(N)$, and $\pm$ stand for the two representations of the quotient
group $\OO(N) / \SO(N) \cong \ZZ_2$ 
given by the determinant. Note that $(Y,+)$ is equivalent to
$(Y,-)$ iff $N$ even and $Y$ has exactly $N/2$ rows. 

Since the completely antisymmetric rank $r$ tensor representation of
$\OO(N)$ whose diagram $Y_r$ consists of a single 
column of height $r$ is equivalent to $\det\otimes Y_{N-r}$, the
representation with diagram $Y$ (such that $r+s\leq N$) is equivalent
to $(Y,+)$ if $r\geq N/2$, and to $(Y',-)$ if $r\leq N/2$,
where $Y'$ arises from $Y$ by replacing the first column of height $r$
by a column of height $N-r$. One easily sees that this relabeling of
the irreducible unitary representations is a bijection.

This proves the analog of Theorem 1.

Turning to the analog of Theorem 2, 
we proceed by exploiting the vanishing 
of the determinant operator (\ref{pol}) in every superselection sector. Since
\bea
X_{11}\cdots X_{N+1,N+1}D_{N+1}^*\ket h =
\Big(\prod_{m=1}^{N+1}2(2h_m-m+1)\Big)\ket h,
\eea
the vanishing of $D_{N+1}$ implies that
\bea
2h_m = m-1 \qquad\hbox{for some positive integer}\quad m\leq N+1.
\eea
In particular, $N':= 2h_\infty$ is a nonnegative integer and $N'\leq
2h_m=m-1\leq N$.   

As a consequence, the representation is realized on the Fock space of $N'$ 
real scalar fields. Assuming that $D_N$ does not vanish, we conclude that 
$N\leq N'$, hence $N'=N$ and $h_\infty =N/2$. 
Convergence of the ground state energy requires $g_i=h_\infty$ up to
an irrelevant finite renormalization,
thus proving the analog of Theorem 2.

\section{Concluding remarks}
Finding all representations of an algebra is often a highly nontrivial
problem. Great progress has been made in the mathematical theory of
highest-weight representations of Lie algebras, and these methods have
been successfully exploited for the classification of unitary positive-energy
representations (superselection sectors) of conformal QFT models in
\emph{two} space-time dimensions.

In \emph{four} space-time dimensions these powerful methods were
thought to be inapplicable, because scalar local quantum fields do not
satisfy commutation relations of Lie type \cite{B76}%
\footnote{There are examples of Poincar\'e covariant local Lie
  fields which violate, however, the spectrum condition; see
  \cite{L67}.}.  
However, {\em bilocal}
quantum fields appearing in certain operator product expansions do
have this property. By virtue of this observation, one can benefit
from the theory of highest-weight modules of Lie algebras in order to
study positive-energy representations in quantum field theory.

This article illustrates the approach on a class of nontrivial
examples, thus building the connection between two important developments in
physics and in mathematics that have taken place unaware of each other
during the last decades.

One is the Doplicher--Haag--Roberts (DHR) theory of superselection sectors in
the framework of algebraic QFT, which establishes the duality between
sectors and gauge symmetry (of the first kind). The other is the
classification of highest-weight unitary modules of certain simple Lie
algebras including $\sp(2n,\RR)$ and $\su(n,n)$. These Lie algebras are
found to be realized in the commutation relations of the simplest
bilocal quantum fields occurring in globally conformal invariant QFT
in $D\geq4$ (even) dimensions.

Obtaining by Lie algebra methods the explicit classification of unitary
positive-energy representations of the commutation relations
satisfied by the bilocal fields, we prove that they are all 
realized in a Fock space representation, corresponding to the 
Segal--Shale--Weil representation in mathematical terminology.

This outcome was expected from the corresponding abstract result
obtained in the DHR theory. However, considerable technical
difficulties are encountered in relating the field representations and
their extensions with the representations of the corresponding
nets. The merit of our study is that it gives an independent
re-derivation of the DHR result directly in the field-theoretic
framework for the special cases at hand. Moreover, we have shown that,
with the canonical choice of the Hamiltonian, the embedded 
{\em Lie algebras} $\uu(\infty,\infty)$ and $\sp(\infty,\RR)$ possess 
the same unitary positive-energy representations as the associative
{\em field algebras}. 

On the other hand, our result facilitates the program of classifying
globally conformal invariant quantum field theories in four
dimensions, because it indicates that without loss of generality
one can ``decouple'' scalar free fields from a model \cite{NRT05}.

\bigskip
\bigskip

{\bf Acknowledgments.} B.B., N.N. and I.T. thank for hospitality the
Institut f\"ur Theoretische Physik der Universit\"at G\"ottingen, where this
work was done. B.B. was supported in part by an FRPD grant from North
Carolina State University. The work of N.N. and I.T. in G\"ottingen was
made possible by an Alexander von Humboldt Research Fellowship and an AvH
Research award, respectively, and was supported in part by the
Research Training Network within FP 5 of the European Commission under
contract HPRN-CT-2002-00325 and by the Bulgarian National Council for
Scientific Research under contract PH-1406.

\vskip5mm

\appendix
\section{Mode expansions of local and bilocal fields}\label{AA}
\setcounter{equation}{0}

The conformal Hamiltonian $H$ is a central element of the Lie algebra
$\so(D) \oplus \RR$ of the maximal compact Lie subgroup of the
conformal group $\SO(D,2)$. The eigenfunctions of the
conformal Hamiltonian form a basis of test functions over the
\emph{compactified} Minkowski space $\M \cong (\SS^{D-1} \times
\SS^1) / \ZZ_2$, and each eigenspace is a finite-dimensional
representation of $\SO(D)$. In the complex parameterization
\cite{T86,N05,NT05} of $\M$ given by%
\footnote{%
The embedding of the Minkowski space $M$ in $\M$ reads 
$z^j = \frac{2x^j}{1+x^2-2ix^0}$
($j=1,\dots,D-1$) and $z^D = \frac{1-x^2}{1+x^2-2ix^0}$\,,
where $x=(x^0,x^1,\dots,x^{D-1})$ and $x^2 = -(x^0)^2 + (x^1)^2 +
\cdots +(x^{D-1})^2$. This is a conformal map belonging to the
connected complex conformal group.} 
\bea\label{eA.1}
\M \, \cong \, \bigl\{\z \in \CC^D : \z = e^{i\tau} \u,\,
\tau \in \RR,\, \u \in\SS^{D-1} \bigr\}
\eea
($\Rightarrow$ $\z^2$ $=$ $(z^1)^2+\cdots+(z^D)^2$ $=$ $e^{2i\tau}$),
these eigenfunctions are the Fourier polynomials
\bea
f_{n,\ell,\mu}(\z) = (\z^2)^n \, h_{\ell,\mu} (\z)  =
e^{i (2n+\ell) \tau} \, h_{\ell,\mu} (\u)\, .
\eea
Here $n$ is an arbitrary integer and $\{h_{\ell,\mu} (\u)\}_{\mu =
  1}^{\har_\ell}$ is a (real) basis of \emph{spherical harmonics} on
$\SS^{D-1}$, i.e., homogeneous harmonic polynomials of degree $\ell =
0, 1$, $\ldots$. The number $\har_\ell$ of spherical harmonics of
degree $\ell$ in $D$-dimensional spacetime equals 
\bea\label{harell}
\har_\ell = \frac{D-2+2\ell}{D-2+\ell}\left({D-2+\ell\atop D-2}\right).
\eea
The $\z$-parameterization of $\M$ is conformally equivalent
to the affine pa\-ra\-met\-ri\-za\-tion of the Minkowski space, and any GCI
(poly)local field $\phi (x)$ can be transformed to a conformally
covariant field in the $\z$-coordinates \cite{N05,NT05}. We introduce
a system of modes $\phi_{n,\ell,\mu}$ of $\phi (\z)$ by
\bea\label{eA.2a}
\phi_{n,\ell,\mu} := \phi [f_{n,\ell,\mu}] =
\frac1{V}
\int\limits_{\M} \phi (\z) f_{n,\ell,\mu}(\z) \, d^D\z ,
\eea 
where $d^{D}\z$ is the (complex) volume form of
$\CC^D$ restricted on the real submanifold $\M$ (this is well-defined
only in even space-time dimension $D$ since otherwise $\M$ is
nonorientable), and $V$ is the (pure imaginary) volume of $\M$.

One can write the collection of all modes of the field $\phi (\z)$
as a formal power series
\bea\label{eA.2}
\phi (\z) \, = \,
\sum_{n \, = \, -\infty}^{\infty} \,
\sum_{\ell \, = \, 0}^{\infty} \,
\sum_{\mu \, = \, 1}^{\har_\ell} \,
f_{n,\ell,\mu}(\z) \, \phi_{-n-\ell-\frac{D}{2},\ell,\mu} \, .
\eea
The complex integral over $\M$ gives rise to a linear functional on
the space of all formal power series of the above type,
called the \emph{residue} \cite[Sect. 3]{BN06} and given explicitly
by
\bea\label{eA.3}
\Res_{\z} \, f_{n,\ell,\mu}(\z) \, := \,
\delta_{n,-\frac{D}{2}} \, \delta_{\ell,0}.
\eea
We can write
\bea\label{eA.4}
\phi_{n,\ell,\mu} \, = \, \Res_{\z} \, \phi (\z) \, f_{n,\ell,\mu}(\z),
\eea
provided we choose $h_{\ell,\mu}(\z)$ 
to be orthonormal with respect to the residue, i.e.,
\bea\label{eA.5}
\Res_{\z} \, (\z^2)^{-\frac{D}{2}-\ell} \, h_{\ell,\mu} (\z) \,
h_{\ell',\mu'} (\z) \, = \, \delta_{\ell,\ell'} \, \delta_{\mu,\mu'} \, .
\eea
An important property of the residue is that it is translation invariant:
\bea\label{eA.transl}
\Res_\z \, \partial_{z^{\alpha}}  f(\z) = 0,
\qquad \alpha=1,\dots,D.
\eea
In addition, it satisfies the Cauchy formula \cite{BN06}
\bea\label{eA.cauchy}
\Res_{\z} \bigl((\z-\w)^2\bigr)^{-\frac{D}{2}}_+ f(\z) = f(\w)
\qquad\text{for}\quad 
f(\z) \in \CC[\![\z]\!],
\eea
where $((\z-\w)^2)^{n}_+$ denotes the formal series resulting
from the Taylor expansion of $((\z-\w)^2)^{n}$ in $\w$ around $0$.

It is possible to characterize GCI fields $\phi (\z)$ as formal power
series of the above type, with properties equivalent to the Wightman
axioms. The corresponding algebraic structure is a higher-dimensional
\emph{vertex algebra} \cite{N05,NT05,BN06}.

For massless scalar fields of canonical scaling dimension
$d_0=(D-2)/2$ in even space-time dimension $D$, only the modes
with $n=0$ and $n=-\ell-d_0$ contribute in
(\ref{eA.2}), which correspond to solutions of the wave equation. The
mode expansion for a pair of conjugate fields then can be conveniently
written as (cf.\ \cite{NT05}):
\begin{align}
\nonumber
\varphi (\z) \, &= \,
\sum_{\ell \, = \, 0}^{\infty} \,
\sum_{\mu \, = \, 1}^{\har_\ell} \,
\Bigl\{
(\z^2)^{-\ell-d_0} \, \varphi_{\ell+d_0,\,\mu}
+
\varphi_{-\ell-d_0,\,\mu}
\Bigr\} \, h_{\ell,\mu} (\z),
\\ \label{eA.6}
\varphi^{*} (\z) \, &= \,
\sum_{\ell \, = \, 0}^{\infty} \,
\sum_{\mu \, = \, 1}^{\har_\ell} \,
\Bigl\{
(\z^2)^{-\ell-d_0} \, \varphi_{\ell+d_0,\,\mu}^{*}
+
\varphi_{-\ell-d_0,\,\mu}^{*}
\Bigr\} \, h_{\ell,\mu} (\z),
\end{align}
where the modes $\varphi_{\pm \ell,\mu}^{(*)}$ are conjugate to each other:
\bea\label{eA.7}
(\varphi_{\ell,\mu})^* \, = \,
\varphi_{-\ell,\mu}^* \, .
\eea
This corresponds to the conjugation law $(\varphi
(\overline{\z}))^* = (\z^2)^{-d_0}\,\varphi^{*} (\z/\z^2)$
reflecting the fact that we
work in a complex parameterization of the real compactified Minkowski space.

In terms of the modes $\varphi_{\ell,\mu}^{(*)}$, the canonical
commutation relations 
\bea\label{eA.8a}
\bigl[ \varphi (\z), \varphi^* (\w) \bigr]
= \bigl((\z-\w)^2\bigr)^{-d_0}_+ -
\bigl((\w-\z)^2\bigr)^{-d_0}_+ 
\eea
become 
\bea\label{eA.8}
\bigl[ \varphi_{\ell+d_0,\mu} \,, \varphi^*_{-\ell'-d_0,\mu'} \bigr] \, = \,
\frac{d_0}{\ell+d_0} \, \delta_{\ell,\ell'} \delta_{\mu,\mu'} \, = \,
\bigl[ \varphi^*_{\ell+d_0,\mu} \,, \varphi_{-\ell'-d_0,\mu'} \bigr] \,,
\eea
and all other commutators vanish. This follows from \eqref{eA.6}
and the orthogonal
harmonic decomposition of $((\z-\w)^2)_+^{-d_0}$ 
\cite[Sect. 3.3]{BN06}: 
\bea\label{eA.8b}
((\z-\w)^2)_+^{-d_0} = \sum_{\ell \, = \, 0}^{\infty} \,
\sum_{\mu \, = \, 1}^{\har_\ell} \,
\frac{d_0}{\ell+d_0} \, (\z^2)^{-\ell-d_0} \, 
h_{\ell,\mu} (\z) h_{\ell,\mu} (\w) \,.
\eea

Choosing an enumeration, $n=n(\ell,\mu) \in \NN$, we define an infinite
number of creation and annihilation operators $a_n^{(*)}$ and
$b_n^{(*)}$ 
for states of positive and negative charge by setting 
\begin{align}\label{eA.9}
a_n &= \sqrt{\frac{\ell+d_0}{d_0}} \, \varphi_{\ell+d_0,\mu} \, ,
\qquad\;\;
b_n = \sqrt{\frac{\ell+d_0}{d_0}} \, \varphi^*_{\ell+d_0,\mu} \, ,
\nonumber \\
a_n^* &= \sqrt{\frac{\ell+d_0}{d_0}} \, \varphi^*_{-\ell-d_0,\mu} \, ,
\qquad
b_n^* = \sqrt{\frac{\ell+d_0}{d_0}} \, \varphi_{-\ell-d_0,\mu} \,.
\end{align}
These operators satisfy the canonical commutation relations
\bea\label{eA.10}
\bigl[a_m,a_n^*\bigr] \, = \, \delta_{m,n} \, = \, \bigl[b_m,b_n^*\bigr],
\quad \bigl[a_m,b_n^{(*)}\bigr] \, = \, 0,\; \text{etc.}
\eea
The conformal Hamiltonian $H$ and the charge operator $Q$ are 
then expressed as
\bea\label{eA.11}
H \, = \, \sum\limits_{n \, = \, 1}^{\infty} \eps_n \,
\bigl(a_n^* a_n + b_n^* b_n \bigr)
\, , \qquad
Q \, = \, \sum\limits_{n \, = \, 1}^{\infty}
\bigl( a_n^* a_n - b_n^* b_n \bigr)
\eea
where $\eps_{n(\ell,\mu)}$ $:=$ $\ell + d_0$ ($\ell = 0,1,\ldots$) 
are the energy eigenvalues.

Introducing the notation 
$h_{n (\ell,\mu)} (\z)$ $:=$ $\sqrt{\frac{d_0}{\ell+d_0}}$ $h_{\ell,\mu} (\z)$,
we rewrite \eqref{eA.6} as follows:
\begin{align}
\nonumber
\varphi (\z) \, &= \,
\sum_{n \, = \, 1}^{\infty} \,
h_n(\z) \, \Bigl\{
(\z^2)^{-\eps_n} \, a_n + b_n^* \Bigr\} \,,
\\ \label{eA.6a}
\varphi^{*} (\z) \, &= \,
\sum_{n \, = \, 1}^{\infty} \,
h_n(\z) \, \Bigl\{
a_n^* + (\z^2)^{-\eps_n} \, b_n \Bigr\} \,.
\end{align}
Similarly, one can write the mode expansion of a complex bilocal
field $W(\z_1,\z_2)$ satisfying the commutation relations (\ref{ccomm}) as
\bea\label{eA.12}
&&
W (\z_1,\z_2) =
\sum_{n_1,n_2 \, = \, 1}^{\infty} h_{n_1} (\z_1) \, h_{n_2} (\z_2)\,
\Bigl\{
X_{n_1 n_2}
+
(\z_1^2)^{-\eps_{n_1}} \, (\z_2^2)^{-\eps_{n_2}} \, X^*_{n_1n_2} 
\nonumber \\ && \quad
+ \,
(\z_1^2)^{-\eps_{n_1}} \,
\Bigl( E_{n_2n_1}^- - \frac{N}{2} \, \delta_{n_1,n_2} \Bigr)
+ \,
(\z_2^2)^{-\eps_{n_2}} \,
\Bigl( E_{n_1n_2}^+ - \frac{N}{2} \, \delta_{n_1,n_2} \Bigr)
\Bigr\} \,
\raisebox{-7pt}{}
\, ,
\eea
where $X^{(*)}_{mn}$ and $E^\pm_{mn} = (E^{\pm}_{nm})^*$ 
satisfy the commutation relations (\ref{ucomm}).

The mode expansion of a real bilocal field $V (\z_1,\z_2)$ satisfying
(\ref{rcomm}) looks exactly the same, but without the superscripts
$\pm$ on $E_{nm}$ and with the symmetry
$X_{mn}^{(*)} = X_{nm}^{(*)}$.

\section{Stress-energy tensor and the conformal Lie algebra}\label{AB}
\setcounter{equation}{0}

The \emph{stress-energy tensor} in any conformal field theory is expected to 
be a local tensor field that gives rise to the space-time symmetry 
generators when integrated against certain functions suggested by the
(classical) Lagrangian field theory.
These integrals are usually ill defined in general axiomatic QFT.
But in the presence of GCI the theory can be extended to the 
compactified Minkowski space as we stated in the previous appendix.
Then one can introduce rigorously the notion of a stress-energy tensor 
without any further assumptions.

We shall formulate the notion of a stress-energy tensor in a GCI QFT 
directly in the $\z$-picture introduced in the previous appendix.
It is a \emph{symmetric tensor} field
$T_{\alpha\beta} (\z)$ $=$ $T_{\beta\alpha} (\z)$,
which is \emph{traceless}:
$T_{\alpha\alpha} (\z)$
$=$ $0$,
and \emph{conserved}:
$\partial_{z^{\alpha}} \, T_{\alpha\beta} (\z)$ $=$ $0$
(summation over repeated indices).
It is assumed also to be a \emph{quasiprimary} 
tensor field of a scaling dimension
equal to the space-time dimension $D$.
These assumptions can be conveniently reformulated using
the following generating function of
$T_{\alpha\beta}$:
\bea\label{B.1}
T (\z;\v) \, = \, T_{\alpha\beta} (\z) v^{\alpha} v^{\beta} .
\eea
Note that $T (\z;\v)$ is a quadratic polynomial in $\v$ with coefficients
that are operator-valued (formal) distributions.

Then, the above postulates for $T_{\alpha\beta}$ read as follows:
\bea\label{B.2-1}
\partial_{\v}^2 \, T (\z;\v) \, = && \hspace{-15pt} 0 \quad
(\text{tracelessness}),
\\ \label{B.3-1}
\partial_{\z} \cdot \partial_{\v} \, T (\z;\v) \, = && \hspace{-15pt} 0 \quad
(\text{conservation law}) \, .
\eea
The statement that $T_{\alpha\beta}$ is a quasiprimary tensor field reads:
\bea
\label{B.2}
\left[ \hspace{1pt} T_{\alpha} \hspace{1pt} , \hspace{1pt}
T(\z;\v) \, \right]
= \! && \hspace{-15pt}
\partial_{z^{\alpha}} \, T(\z;\v)
\, , \quad
\raisebox{9pt}{}
\\ \label{B.3}
\left[ \hspace{1pt} H \hspace{1pt} , \hspace{1pt}
T(\z;\v) \, \right]
= \! && \hspace{-15pt}
(\z \cdot \partial_{\z} + D) \, T(\z;\v)
\, , \quad
\raisebox{9pt}{}
\\ \label{B.4}
\left[ \hspace{1pt} \Omega_{\alpha\beta} \hspace{1pt} , \hspace{1pt}
T(\z;\v) \, \right]
= \! && \hspace{-15pt}
\bigl( z^{\alpha} \, \partial_{z^{\beta}} - z^{\beta} \, \partial_{z^{\alpha}}
+ v^{\alpha} \, \partial_{v^{\beta}} - v^{\beta} \, \partial_{v^{\alpha}}
\bigr) \, T(\z;\v) \,, \quad
\\ \label{B.6}
\left[ \hspace{1pt} C_{\alpha} \hspace{1pt} , \hspace{1pt}
T(\z;\v) \, \right]
= \! && \hspace{-15pt}
\bigl(
\z^2 \, \partial_{z^{\alpha}} - 2 \, z^{\alpha} \, \z \cdot \partial_{\z} - 
2D \, z^{\alpha} \nonumber \raisebox{11pt}{} \\ && \hspace{-15pt}
+ 2 \, \z \cdot \v \, \partial_{v^\alpha} - 2 \, v^\alpha \, \z \cdot 
\partial_\v \bigr) \, T(\z;\v) \,,
\eea
where $T_\alpha$, $H$, $\Omega_{\alpha\beta} = -\Omega_{\beta\alpha}$ 
and $C_\alpha$ 
are the generators of the conformal Lie algebra $\so(D,2)$,
which satisfy the relations:
\bea\label{e1.1}
\hspace{-1.14pt}
\left[ H , \Omega_{\alpha\beta} \right]
\hspace{-1pt} = \hspace{-15pt} && \hspace{-2pt} 0 = \hspace{-1pt}
\left[ T_{\alpha} , T_{\beta} \right]
\hspace{-1pt} = \hspace{-1pt}
\left[ C_{\alpha} , C_{\beta} \, \right]
\, ,
\nonumber \\ \hspace{-1.14pt}
\left[ \Omega_{\alpha_1\beta_1} ,
\Omega_{\alpha_2\beta_2} \right]
\hspace{-1pt} = \hspace{-15pt} && \hspace{-2pt}
\delta_{\alpha_1\alpha_2}\,\Omega_{\beta_1\beta_2}
\! +
\delta_{\beta_1\beta_2}\,\Omega_{\alpha_1\alpha_2}
\! -
\delta_{\alpha_1\beta_2}\,\Omega_{\beta_1\alpha_2}
\! -
\delta_{\beta_1\alpha_2}\,\Omega_{\alpha_1\beta_2}
,
\raisebox{10pt}{} \nonumber \\ \hspace{-1.14pt} &&
\hspace{-51pt}
\hspace{-15.5pt} \hspace{8pt} \hspace{-3.53pt} \hspace{-2pt}
\begin{array}{rlrl}
\left[ H , T_{\alpha} \, \right]
\hspace{-1pt} = \hspace{0pt} & \hspace{-8pt}
T_{\alpha}
\, , \
&
\left[ H , C_{\alpha} \, \right]
\hspace{-1pt} = \hspace{0pt} & \hspace{-8pt}
- C_{\alpha}
,
\raisebox{10pt}{} \nonumber \\
\left[ \Omega_{\alpha\beta} ,
T_{\gamma} \, \right]
\hspace{-1pt} = \hspace{0pt} & \hspace{-8pt}
\delta_{\alpha\gamma} T_{\beta}
\! - \!
\delta_{\beta\gamma} T_{\alpha}
,
& \ \
\left[ \Omega_{\alpha\beta} ,
C_{\gamma} \right]
\hspace{-1pt} = \hspace{0pt} & \hspace{-8pt}
\delta_{\alpha\gamma} C_{\beta}
\! - \!
\delta_{\beta\gamma} C_{\alpha}
,
\raisebox{12pt}{}
\end{array}
\raisebox{14pt}{} \nonumber \\ \hspace{-1.14pt}
\left[ T_{\alpha} , C_{\beta} \right]
\hspace{-1pt} = \hspace{-15pt} && \hspace{-2pt}
2 \, \delta_{\alpha\beta} H
- 2 \, \Omega_{\alpha\beta} \,.
\eea

Since $T_{\alpha\beta}$ is a tensor field it requires 
``tensor test functions''. For
\bea\label{B.5-1}
f (\z;\v) \, = \, f_{\alpha\beta} (\z) \, 
v^{\alpha} v^{\beta}
\, , \quad f_{\alpha\beta} (\z) \, \in \, \CC [\z,1/\z^2]
\,,
\eea
we define
\bea\label{B.4-1}
T [f] \, := \, \frac{1}{2} \ \Res_\z \,
\Bigl\{
f (\z,\partial_{\v}) \, T (\z;\v)
\Bigr\}
\,.
\eea
Using the residue technique of \cite{BN06} (see Appendix~\ref{AA}),
one can derive the following statement.

\medskip

\noindent
\textbf{Proposition 1.}
{\it
Let $T (\z;\v)$ be a local field
defined by (\ref{B.1}) and satisfying relations
(\ref{B.2-1})--(\ref{B.2})
in a vertex algebra, which is not assumed
to be conformal in advance.
Introduce the operators $X := T [f_X]$ for
$X$ $=$ $T_{\alpha\beta},$ $H,$ $\Omega_{\alpha\beta},$ $C_{\alpha}$
($\alpha,\beta$ $=$ $1,\ldots,D$), where
\bea\label{BX}
&
f_{\text{\tiny $X$}} \, = \,
{\displaystyle \frac{\z \cdot \v}{\z^2}} \ g_{\text{\tiny $X$}} (\z,\v) \,,
\qquad
g_{\text{\tiny $T_{\alpha}$}} \, = \, v^{\alpha}
\, , \quad
g_{\text{\tiny $H$}} \, = \, \v \cdot \z
\, ,
& \nonumber \\ &
g_{\text{\tiny $\Omega_{\alpha\beta}$}} \, = \,
z^{\alpha} v^{\beta} - z^{\beta} v^{\alpha}
\, , \quad
g_{\text{\tiny $C_{\alpha}$}} \, = \,
\z^2 \,v^{\alpha} - 2 \, z^{\alpha} \, \z \cdot \v
\,.
&
\eea
Then these operators
obey the conformal Lie algebra relations (\ref{e1.1})
if and only if Eqs.~(\ref{B.3})--(\ref{B.6}) hold.}

\medskip

Given the bilocal field $W(\z,\w)$ or $V(\z,\w)$, 
we can define the stress-energy tensor $T (\z;\v)$ by
applying to the bilocal field
a second order differential operator $\mathcal{D}$ $=$
$\mathcal{D} (\partial_{\z},\partial_{\w};\v)$ and equating the arguments 
\cite{NRT05}:
\bea\label{B.12}
T (\z;\v) \, = \,
\mathcal{D} \, W (\z,\w)
\bigl|_{\w \, = \, \z}
\, , \quad
\text{or} \quad
T (\z;\v) \, = \,
\frac{1}{2} \,
\mathcal{D} \, V (\z,\w)
\bigl|_{\w \, = \, \z}
\,,
\eea
where
\bea\label{B.13}
(D-1) \mathcal{D} (\partial_{\z},\partial_{\w};\v) =
d_0
\bigl[
(\v \cdot \partial_{\z})^2 + (\v \cdot \partial_{\w})^2
\bigr]
-
D \, \bigl(\v \cdot \partial_{\z}\bigr) \bigl(\v \cdot \partial_{\w}\bigr)
+
\v^2 \bigl(\partial_{\z} \cdot \partial_{\w}\bigr)
\;
\eea
and $d_0 := (D-2)/2$.
Note that the second formula in (\ref{B.12}) follows from the first for
$V (\z,\w)$ $=$ $W (\z,\w) + W (\w,\z)$.
It is an easy exercise to verify that the harmonicity of $W(\z,\w)$ 
(or of $V(\z,\w)$)
implies both the tracelessness and the conservation of $T (\z;\v)$;
for instance,
\bea\label{B.14}
(D-1)
&& \hspace{-17pt}
\bigl(\partial_{\z}+\partial_{\w}\bigr) \cdot \partial_{\v} \
\mathcal{D} (\partial_{\z},\partial_{\w};\v) \, W (\z,\w)
\nonumber \\ && \hspace{-17pt} \, = \,
- 2 \,
\bigl( (\v \cdot \partial_{\z}) \, \partial_{\z}^2 + (\v \cdot \partial_{\w}) 
\, \partial_{\w}^2 \bigr) \, W (\z,\w) \, = \, 0
\, .
\eea

\medskip

\noindent
\textbf{Proposition 2.}
{\it
Let $W (\z_1,\z_2)$ be given by (\ref{eA.12}).
Then $T (\z;\v)$ defined by (\ref{B.12}) generates a representation of
the conformal Lie algebra by (\ref{BX}).
Furthermore, $W (\z_1,\z_2)$ transforms under this representation as a
scalar bilocal field of dimension $(d_0,d_0)$; in particular,
\bea\label{B.15}
\bigl[ H,W (\z_1,\z_2) \bigr] \, = \,
\bigl(\z_1 \cdot \partial_{\z_1} + \z_2 \cdot \partial_{\z_2} + 2 \, 
d_0\bigr) \, W (\z_1,\z_2) \, .
\eea
A similar statement is valid for $V (\z_1,\z_2)$.
}
\medskip

The proposition shows that, if one wants to realize {\it both}
$\uu(\infty, \infty)$ and $\so(D,2)$ in the state space of the theory, one
cannot absorb the central term in (\ref{ccomm}) involving the constant 
$N$ by a redefinition of the field $W(x_1,x_2) \mapsto W(x_1,x_2) - (N/2)
(\Delta^+_{1,2}+\Delta^+_{2,1})$, without its reappearance
in formula (\ref{B.12}) for the generators of $\so(D,2)$.

\end{document}